\documentclass{aa}
\usepackage{txfonts}
\usepackage{amssymb,epsfig,amscd}
\usepackage{graphicx}% Include figure files
\usepackage{dcolumn}% Align table columns on decimal point
\usepackage{bm}% bold math

\usepackage{epsfig}

\def\be{\begin{equation}} 
\def\ee{\end{equation}}
\def\bq{\begin{eqnarray}} 
\def\eq{\end{eqnarray}}

\begin{document}

\title{Gravitational wave bursts induced by \\ r-mode spin-down of
hybrid stars}

\author{A. Drago\inst{1}, G. Pagliara\inst{1} \and Z. Berezhiani\inst{2}}
\institute{Dipartimento di Fisica, Universit{\`a}
di Ferrara and INFN, Sezione di Ferrara, 44100 Ferrara, Italy \and
Dipartimento di Fisica, Universit\`a di L'Aquila and INFN,
Laboratori Nazionali del Gran Sasso, 67010 L'Aquila, Italy}

%\date{\today } 

\abstract{ 
We show that sudden variations in the composition and structure of an
hybrid star can be triggered by its rapid spin-down, induced by r-mode
instabilities.  The discontinuity of this process is due to the
surface tension between hadronic and quark matter and in particular to
the overpressure needed to nucleate new structures of quark matter in
the mixed phase. The consequent mini-collapses in the star can produce
highly energetic gravitational wave bursts. The possible connection
between the predictions of this model and the burst signal found by
EXPLORER and NAUTILUS detectors during the year 2001 is also investigated.}

%\pacs{04.30.Db, 25.75.Nq, 26.60.+c, 97.60.Jd}

\noindent

\maketitle 


\section{Introduction}

The analysis of the data collected by the Gravitational Wave antennas
EXPLORER and NAUTILUS during the year 2001 shows the existence of
coincidences between the signal detected by these two resonant
bars (\cite{Astone:2002ra,Coccia:2004gw}). In particular, an excess of
coincidences respect to the background is concentrated around sidereal
hour four, which corresponds to the orientation for which the
sensitivity of the bars is maximal for a signal coming from the
direction of the galactic center.  Although the statistical
significance of this signal is debated
(see \cite{Finn:2003bd,Astone:2003ng,Astone:2003ft}), it is interesting to
investigate the possible origin of the inferred signal using existing
models of GW emission. In particular we will concentrate on the ones
associated with instability modes of a rotating compact star.  It is
important to remark that in the analysis performed on the data from
the two gravitational antennas only impulsive signals have been
searched for, having a typical duration of $\sim$ 10 ms.  In the
literature an extensive analysis has been performed concerning steady
sources and periodic GW emission
(\cite{Andersson:2001ev,Wagoner:2002vr,Reisenegger:2003cq}).  In this
paper we will explore the possibility that a transient GW burst is
triggered by the steady emission of GWs generated by r-mode
instabilities.  In particular, we will show that the almost continuous
dragging of angular momentum from the star can induce a sudden
variation in its structure and composition, generating a few bursts of
GWs.

The first ingredient of our model are the so called r-mode
instabilities, which are very efficient in dragging
angular momentum from a rapidly rotating compact star (\cite{Andersson:1998xt,Friedman:1998uh}).
A rapid spin-down 
deeply affects the structure and composition of the star. This problem
has been discussed in detail in a few papers in the past years
(see \cite{Glendenning:1997fy,Chubarian:1999yn}), where it has been shown
that, using rather standard values for the model parameters, the
composition of an hybrid hadronic-quark star can be dramatically
modified when the rotational frequency changes in a range centered
around a few hundreds Hertz. In particular, during the spin-down
era the central density increases and a larger amount of matter is
converted into a mixed hadron-quark phase (MP) or into pure quark
matter. The new idea we discuss in this paper is the effect of a surface tension at the interface
between hadronic and quark matter. If the surface tension is not vanishing, the formation of new
structures in the MP can be delayed. A similar idea has been discussed
recently in connection with the transition from a metastable hadronic
star into a stable hybrid or quark star, showing that a huge amount of
energy can be released in the process of conversion of bulk hadronic
matter into quark matter (\cite{Berezhiani:2002ks,Drago:2004vu,Bombaci:2004mt}) 
\footnote{A similar idea has recently been discussed in Ref.~(\cite{Harko:2004zz}).
There, the possibility that the spin down can trigger a phase transition is
analyzed taking into account the effect of the surface tension. In Ref.~(\cite{Harko:2004zz})
the angular momentum of the star is dragged by the magnetic
dipole radiation which can be an efficient mechanism of spin down only
in presence of a strong magnetic field.}. 
The mechanism discussed in the present paper
is also based on metastability, but it involves phase transitions in
the MP, which are responsible for a relatively large modification of
the stellar structure without a large energy release. We suggest the
possibility that the formation of the MP can take place in several
steps, in each of which the radius and the oblateness of the star
change only by a relatively small amount.  
A crucial observation is that the conversion process from hadronic
matter into MP will start in one point inside a spheroidal layer of
metastable matter (to be discussed later) and will propagate inside
that same layer at finite velocity. In this way the burning process
generates a large non-radial oscillation of the star.
During this period a strong GW
emission takes place, 
and we will discuss its possible connection with
the analysis of the gravitational bars data by the ROG Collaboration~\footnote{
Another possible relation between phase transitions and GW emissions is discussed in
Ref. (\cite{Miniutti:2002bh}). In
that paper phase transitions, associated with density discontinuities,
can excite the so called g-modes and large non-radial oscillations can
develop.}.

\section{R-mode induced spin down}

\begin{figure}[]
\begin{center}
\includegraphics[scale=0.52]{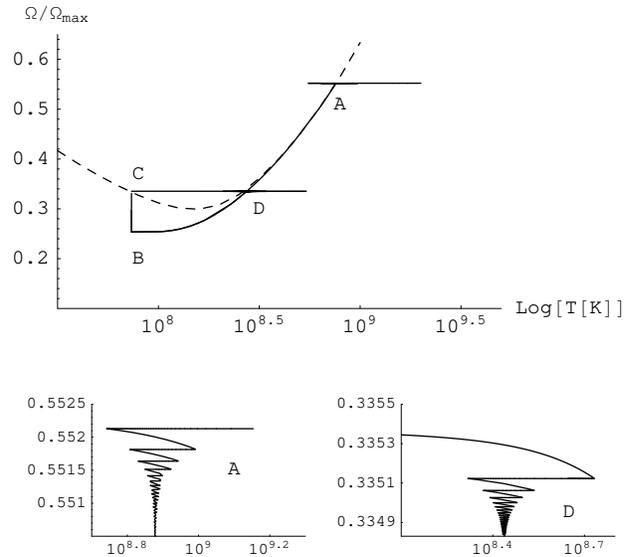}
\end{center}
\parbox{8cm}{
\caption{r-modes instability window (upper panel). Coming 
from high temperatures (A) the star loses its
angular momentum until it exits the instability region (B) (first scenario). By mass accretion
the star can be re-accelerated until it reaches the instability window from the low 
temperature side (C). The excitation of r-modes instability produces a fast
reheating of the star due to bulk viscosity dissipation (D) (second scenario).
In the lower panel a magnification of the regions indicated by (A) and (D) is displayed.
Here we used a mass accretion rate 
of $10^{-8} M_\odot$ per year.  
\label{ciclo}}
}
\end{figure}

In the literature, two scenarios have been discussed
concerning a steady GW emission due to r-mode instabilities. 
The first scenario is based on the emission of
GWs from a hot and rapidly spinning compact stellar object, which has
not lost its angular momentum in the very first part of its existence
after the supernova explosion.  This is possible if the bulk viscosity
is large for temperatures of order $\sim 10^9$K or higher, so that
r-modes are damped till the temperature drops below $\sim 10^9$K.
Possible candidates are quark stars (\cite{Andersson:2001ev}), hyperonic
stars (\cite{Lindblom:2001hd}) or hybrid stars (\cite{Drago:2003wg}).  The
second scenario involves older stars which are reaccelerating due to
mass accretion from a companion. In this way a sort of ``cycle'' can
develop (see Fig.~\ref{ciclo}), in which the star periodically goes through the
following steps: 1) mass accretion with increase of angular velocity;
2) instability due to r-mode excitations with reheating due to bulk
viscosity; 3) loss of angular momentum with emission of GWs.  The reheating 
due to bulk viscosity during phase 2) is so
efficient that, if the instability region is reached from the
low-temperature side, the star rapidly reheats and reaches
the high-temperature side of the instability region
(\cite{Reisenegger:2003cq})~
\footnote{Also neutron stars can recycle, 
as suggested in Refs.~(\cite{Levin:1998wa,Andersson:2000pt,Heyl:2002pe}).}.

%%%%%%%%%%%%%%%%%%%%%%%%%%%%%%%%%%%%%%%%%%%%%%%%%%%%%%%%%%%%%%%%%%%%%%%%%%%%%

%%%%%%%%%%%%%%%%%%%%%%%%%%%%%%%%%%%%%%%%%%%%%%%%%%%%%%%%%%%%%%%%%%%%

\begin{figure}[t]
\begin{center}
\includegraphics[scale=0.52]{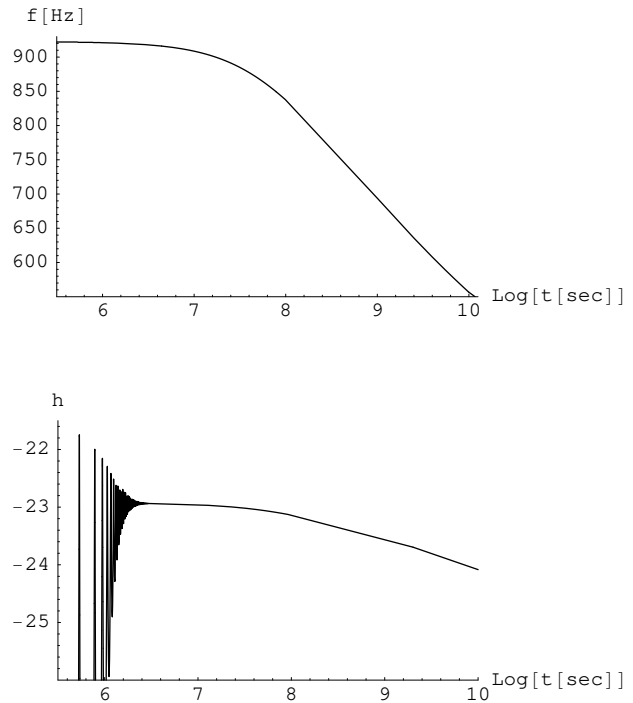}
\end{center}
\parbox{8cm}{
\caption{The frequency (upper panel) and amplitude (lower panel) of the GW signal
emitted by r-mode instabilities as functions of time.
Here the compact star is 
approaching the instability window from the high temperature side (first scenario). 
The computed amplitude corresponds to a distance of 1 kpc. 
\label{sig1}}
}
\end{figure}

\begin{figure}[t]
\begin{center}
\includegraphics[scale=0.52]{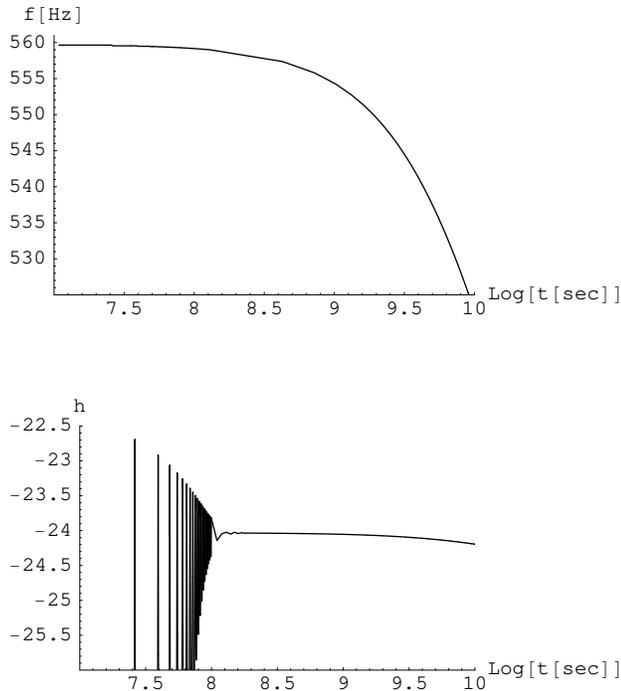}
\end{center}
\parbox{8cm}{
\caption {Same as in fig.~1. Here the
compact star is approaching the instability window from the low temperature side
after the '' recycling'' process (second scenario). 
\label{sig2}}
}
\end{figure}

Let us introduce the equations regulating the process of GW emission
in the two scenarios discussed above. Here we follow the formalism
of \cite{Wagoner:2002vr} and \cite{kinney}, where the
scheme proposed by \cite{owen} has been refined
\footnote{We assume
that the r-mode canonical angular momentum contributes zero
physical angular momentum to the star. This is a sensible approximation
as results both from explicit calculations (\cite{Levin2}) and also
from the observation that
the r-modes saturate quickly and after saturation no physical angular
momentum can go into them (\cite{Wagoner:2002vr,kinney}).}.
%, in particular, the expression for
%the evolution of the angular momentum of the star is needed.  In a
%first study on this topic (\cite{owen}), it was assumed that the total
%angular momentum of the star is the sum of the equilibrium angular
%momentum and the canonical angular momentum of the r-mode. As observed
%in Ref.~(\cite{Levin2}), the physical angular momentum of the r-mode
%does not coincide with its canonical angular momentum and therefore
%the evolution equations of Ref.~ (\cite{owen}) are not
%correct. Actually, as assumed in Ref.~(\cite{Wagoner:2002vr}), the
%physical angular momentum of the r-mode can be written as the product
%of the canonical angular momentum and a numerical coefficient
%$(1-K_j)$. Here, following Ref.~(\cite{kinney}), we take $K_j=1$
%meaning that the r-modes canonical angular momentum contributes zero
%physical angular momentum to the star.  The exact value of $K_j$ is
%not important as remarked in Refs.~(\cite{Wagoner:2002vr,kinney})
%since the r-modes saturate quickly and no more physical angular
%momentum can go into them.  Finally, the evolution equations read:
The evolution equations read:
\bq
\noindent&&{\dot{\alpha}\over\alpha}=
{-1\over  t_v}-\left(1+{3 \alpha^2\tilde J\over 2 \tilde I} \right)
\left({1\over  t_g}\right)-{\dot{M}\over 2\tilde I \Omega}
\left({G\over M R^3}\right)^{1\over 2}\\
&&\dot\Omega={\dot M\over\tilde I}\left({G\over MR^3}\right)^{1/2}-
{\dot M\Omega\over M}+3\Omega\alpha^2 {\tilde J\over\tilde I}\left({1\over t_{g}}\right)\\
&&\dot E_{\mathrm{thermal}}=\dot E_{\mathrm{accretion}}+\dot E_{\mathrm{viscosity}}
-\dot E_{\mathrm{neutrino}}\, .
\eq
Here $\alpha$ is the dimensionless amplitude of the r-mode,
$1/t_v \equiv 1/t_s +1/t_b $,  $t_g$, $ t_s$ and $t_b$ are time scales associated 
with GW emission, to shear
and to bulk viscosity damping, respectively.  $\tilde I$ and $\tilde
J$ are dimensionless values of the moment of inertia and of the
angular momentum (for all details see
e.g. Ref.~(\cite{Andersson:2001ev})). Eq.~(1) describes the damping of
r-modes due to viscosity, Eq.~(2) describes
the variation of the angular momentum of the star in presence
of the torque due to gravitational wave emission and to
mass accretion and, finally, Eq.~(3) describes the thermal evolution
given by the contributions of the reheating due to mass accretion,
shear and bulk viscous dissipation of the r-modes and of the cooling due to
neutrino emission. Obviously, in the first scenario depicted above,
mass accretion is not present.  To compute the time scale
associated with bulk viscosity for hybrid stars we use the results of
Ref.~(\cite{Drago:2003wg}) in which the viscosity of MP has been computed.
In that paper it has been shown that the viscosity of MP is of the  
same order of magnitude as the viscosity of pure quark matter if superconducting gaps are not present 
or it is reduced by a factor 
$\sim$ 10 if a color superconducting 2SC gap is taken into account~(see \cite{Madsen:1999ci}).
Concerning the value of shear viscosity, we have taken into account 
not only the contribution associated with pure quark matter (\cite{Andersson:2001ev}), but 
also the contribution associated with the 
viscous boundary layer which is present in a star having a crust made of nucleonic matter
(\cite{Bildsten:1999zn}).
In Fig.~\ref{sig1} we plot the dimensionless amplitude $h$ and the frequency $f$ of the GWs emitted by a 
a star in the first scenario.
In Fig.~\ref{sig2}, the same quantities are displayed for the
second scenario.
In this figure, a mass accretion rate of order $10^{-8} M_\odot$ per year
has been assumed. If the value of the accretion rate is reduced, the star re-enters the
instability window at smaller temperatures and larger values of the angular velocity. In this way,
the GW emission in the second scenario would be rather similar to the one obtained in the first scenario.
Finally, let us remark that the same compact star can enter first the
instability window associated with the first scenario, when the star is relatively young,
and, after some time, the star becomes again unstable
due to mass accretion as described in the second scenario. These possibilities are displayed in Fig.~\ref{ciclo}.

As it can be seen in Figs.~\ref{sig1}--\ref{sig2}, in both scenarios the initial part of the
signal decomposes into bursts lasting a few minutes and separated by 
periods of a few days (or months) of
quiescence. During this initial phase of the emission, the star follows a trajectory in the
temperature -- angular velocity plane, oscillating around the
instability line displayed in Fig.~\ref{ciclo} (see also Ref.~(\cite{Andersson:2001ev})).  After
this phase, which can last months or years, the angular momentum is
dragged almost continuously and the signal becomes steady for hundreds
of years until the star finally exits the instability region.  In
principle, the first part of the signal having an amplitude h $\sim 10^{-22}$ for a source
located at 1 kpc, could be detected by 
resonant bars. The main difficulty associated with the search of this type of
periodic signal is due to the almost monocromaticity of the GW emission which could be detected
only by a dedicated search. For instance, in the case of the pulsar J1939+2134, a dedicated search
has been performed for a signal having a frequency twice the rotation frequency of the 
star (which would correspond to the signal emitted by a stellar object having a non-axisimmetric shape), 
with a null result (see \cite{Abbott:2003yq}). On the other hand, the signal associated with r-modes would 
mainly be emitted
at a frequency $f_g=2\Omega/3\pi$.

%%%%%%%%%%%%%%%%%%%%%%%%%%%%%%%%%%%%%%%%%%%%%%%%%%%%%%%%%%%%%%%%%%%%%

%%%%%%%%%%%%%%%%%%%%%%%%%%%%%%%%%%%%%%%%%%%%%%%%%%%%%%%%%%%%%%%%%%%%

%%%%%%%%%%%%%%%%%%%%%%%%%%%%%%%%%%%%%%%%%%%%%%%%%%%%%%%%%%%%%%%%%%%%

\section{Toy-model for a rotating star \label{sectoy}}

The r-mode mechanism we have discussed so far produces a steady
emission of periodic GWs and therefore it does not correspond to the
signal discussed in Refs.~(\cite{Astone:2002ra,Coccia:2004gw}).  The new
idea we are now introducing is a possible connection between the rapid
dragging of the angular momentum and discontinuous changes of the
structure of the star.  It is worth remarking that r-modes are a very
efficient way of reducing the angular momentum of a compact
star. Although it will not be discussed here, also a strong magnetic
field as the one present in magnetars can efficiently dragg the angular momentum
of the star by dipole magnetic radiation (\cite{Harko:2004zz}). As shown in
Refs.~(\cite{Glendenning:1997fy,Chubarian:1999yn}), when a hybrid star
slows down its density increases till the first critical density is
reached at the center of the star and MP starts being produced. A
further reduction of the angular momentum allows the star to increase
the fraction of MP occupying its core, until the central density
reaches the second critical value. At that point pure quark matter
starts being produced.  During all the spin-down process the radius of
the star decreases by several kilometers.  This gradual modification
of the structure of the star can take place discontinuously in several
(small) steps if the effect of a non-vanishing surface tension is
taken into account. The inner regions of the star in which MP will
later form becomes firstly metastable. The formation of germs of
stable MP proceeds through quantum tunneling. The probability of this
transition strongly depends on the value of overpressure.  As shown in
Refs.~(\cite{Iida:1998pi,Berezhiani:2002ks,Drago:2004vu,Bombaci:2004mt})
the nucleation time needed to form new structures of quark matter can
be very long if the overpressure is not large. When the overpressure
reaches the critical value, in one randomly chosen site inside the
metastable layer a new drop of quark matter forms.  As it will be
shown, the process of conversion of hadronic into quark matter
propagates with finite velocity $v_c$ inside the star and a sudden
modification of the composition and of the structure of the star
occurs during a timescale $\sim R/v_c$.  During this period non-radial
modes develops and a few bursts of GWs can be emitted until a new
equilibrium configuration is reached.

In order to give a qualitative estimate of the magnitude of the oscillations and 
therefore of the amplitude of the GWs emitted we resort to the
toy model proposed in Ref.~(\cite{Heiselberg:1998vh}). 
We model the hybrid star  
as a spheroid containing a MP core with uniform
density $\rho_2$ and a crust of nuclear matter with uniform
density $\rho_1$. The pressure at which the phase transition from the first to the 
second component takes place is a parameter $P_0$.
The Newtonian hydrostatic equation reads:
\be
\frac{1}{\rho}\frac{dP}{da} = -G\frac{m(a)}{a^2} + \frac{2}{3}\Omega^2 a \, ,
\ee
where
\be
m(a) = 4\pi\int^a_0\rho(a')a'^2da'\, .
\ee
Here the ``effective radius'' $a$ appears, which is related to a position
inside the star by the equation:
\be 
r=a  (1-\epsilon  P_2(cos \theta))\, ,\label{radeff}
\ee
where $\epsilon = \frac{5}{4} \frac{\Omega^2}{2 \pi G \rho1}$ and 
$P_2(cos \theta)$ is the 2nd Legendre polynomial.
Eq.~(4) is then solved and in this way the pressure inside the star is
analytically determined for a given angular velocity.  In Fig.~\ref{raggi} the
structure of a rotating star is displayed. Comparing with results
based on the actual solution of the equilibrium equation for a
rotating compact object (see
e.g. (\cite{Glendenning:1997fy,Chubarian:1999yn})), it is possible to
notice that the toy model is at least qualitatively in agreement with
the exact solution.  Unfortunately, the toy model badly underestimates
the quantitative effect of the spin-down on the radius and composition
of the star. For instance, in the toy model we get reductions of the
radius of order 0.5 km, while the calculations based on realistic
equations of state give reductions typically of order a few km
(\cite{Glendenning:1997fy,Chubarian:1999yn}). The same problem arises
for the estimate of the fraction of the volume of the star occupied by
MP: the toy model gives an increase of this quantity of roughly
20$\%$, while more realistic calculations give a significantly larger
value.  It is clear that we are at the moment only aiming at a
qualitative study of the proposed mechanism and that refined
calculations are needed in order to get a more precise estimate.

\begin{figure}[b]
\begin{center}
\includegraphics[scale=0.52]{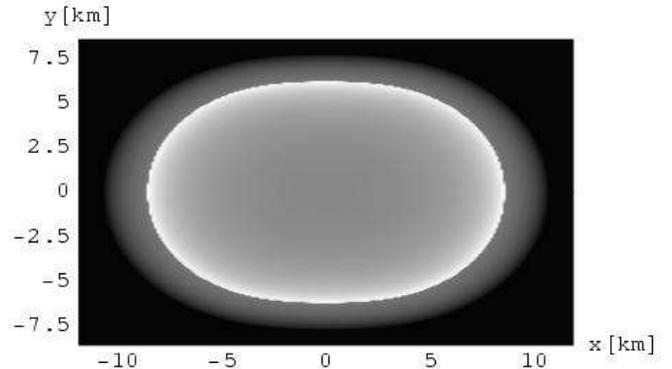}
\end{center}
\parbox{8cm}{
\caption{The value of the overpressure inside the star, for 
$\Omega = 5000 \, s^{-1}$
and $\Delta \Omega/\Omega =0.05$. Lighter areas correspond to higher values of 
the overpressure.
\label{sovrapressione}}
}

\end{figure}

The question we want to numerically investigate within the toy model
is the following: given a specific equation of state (in our case
fixing the values of $\rho_1$, $\rho_2$ and $P_0$) and a value for the
surface tension $\sigma$, which variation of the angular velocity
$\Delta \Omega/\Omega$ is large enough to trigger the formation of a
critical drop of quark matter, in a time scale of order days or years?
The crucial ingredient in this calculation is the relation between the
overpressure $\Delta P/P$ and $\Delta \Omega/\Omega$.  The
overpressure is determined computing, for a same element of fluid, the
difference between the value of the pressure after and before the slow
down (lagrangian perturbation).  In particular, we are interested in
the value of the overpressure in the region immediately surrounding
the core of already formed MP. In Fig.~\ref{over} we show the overpressure, in
the above defined region, for a value $\Delta \Omega/\Omega = 0.05$
and for various values of $\Omega$. As it appears, $\Delta P/P\sim
\alpha\,\,\Delta \Omega/\Omega $, with $\alpha\sim 0.3\div 1.4$, and
the larger values correspond to the faster rotating stars.  It is also
interesting to notice that, as the star slows down the layer in which
new MP is formed moves to outer regions and the thickness of the layer
shrinks. In Fig.~\ref{sovrapressione} the overpressure inside the star
is displayed for a fixed value of $\Omega$ and  $\Delta \Omega/\Omega$.
Notice that the value of the overpressure is larger in the 
layer around the MP core and therefore in that region the 
conversion will take place with a larger probability.

We need now to compute the nucleation time for the obtained value of
the overpressure and a given value of $\sigma$. This can be done
following the formalism developed in
Refs.~(\cite{Iida:1998pi,Berezhiani:2002ks}) and based on quantum
tunneling. Within the toy model, nucleation time of order days can be
obtained, for the values of the overpressure as the one displayed in
Fig.~\ref{over} and using values of $\sigma\sim$ a few MeV/fm$^2$.  These
values of $\sigma$ are of the same order of the one estimated in the
MIT bag model (\cite{Berger:1987ps}). They are also not far from the
values investigated in
Refs.~(\cite{Berezhiani:2002ks,Drago:2004vu,Bombaci:2004mt}).  We would
like to stress again that refined calculations are needed in order to
get a quantitative estimate of the nucleation time.  The most
artificial feature of the toy model we are using is probably related
to the absence of modifications of the already formed MP during the
spin-down process. In a realistic model, not only the volume occupied
by MP increases during the spin-down, but also new structures form,
with an evolution of the existing structures from drops, to rods, to
slabs as the star slows down (\cite{Glend:01kn}).

\section{Emission of GW bursts}

\subsection{Time structure of the bursts}

%%%%%%%%%%%%%%%%%%%%%%%%%%%%%%%%%%%%%%%%%%%%%%%%%%%%%%%%%%%%%%%%%%%%

\begin{figure}[]
\begin{center}
\includegraphics[scale=0.52]{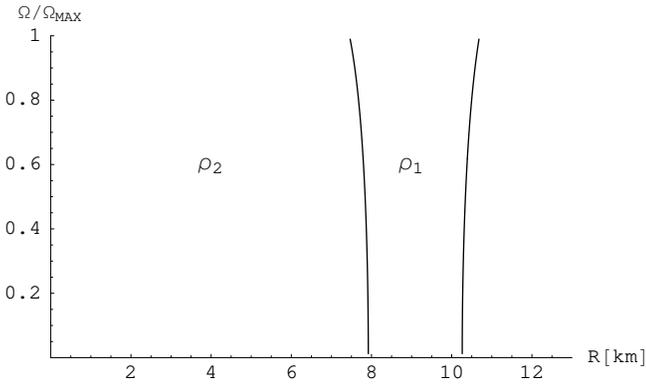}
\end{center}
\parbox{8cm}{
\caption{The total radius of the star and the radius of the inner core of MP are shown 
as functions of the angular velocity in the toy model. 
\label{raggi}}
}
\end{figure}

%%%%%%%%%%%%%%%%%%%%%%%%%%%%%%%%%%%%%%%%%%%%%%%%%%%%%%%%%%%%%%%%%%%%
%%%%%%%%%%%%%%%%%%%%%%%%%%%%%%%%%%%%%%%%%%%%%%%%%%%%%%%%%%%%%%%%%%%%

\begin{figure}[]
\begin{center}
\includegraphics[scale=0.52]{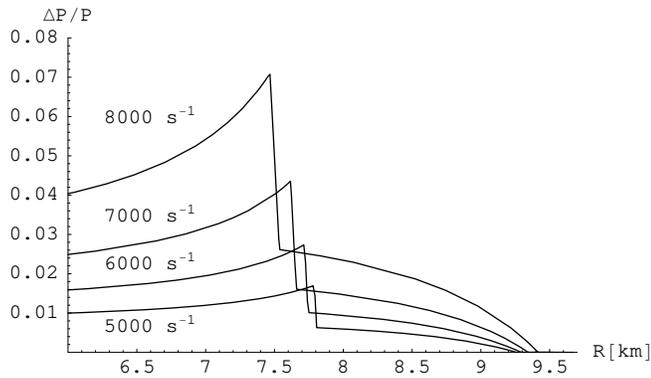}
\end{center}
\parbox{8cm}{ 
\caption {The value of the overpressure as a function of the effective radius, for four values of 
the angular velocity. 
Here $\Delta \Omega/\Omega =0.05$. 
\label{over}}
}
\end{figure}

%%%%%%%%%%%%%%%%%%%%%%%%%%%%%%%%%%%%%%%%%%%%%%%%%%%%%%%%%%%%%%%%%%%%

%%%%%%%%%%%%%%%%%%%%%%%%%%%%%%%%%%%%%%%%%%%%%%%%%%%%%%%%%%%%%%%%%%%%%%%%%%%%%%%%%%%%%%%%%%%%%%%%%%%%%%%%%5 
%%%%%%%%%%%%%%%%%%%%%%%%%%%%%%%%%%%%%%%%%%%%%%%%%%%%%%%%%%%%%%%%%%%%%%%%%%%%%%%%%%%%%%%%%%%%%%%%%%%%%%%%%%%%%%%%%%%%

We can now describe more precisely the GW emission in the mechanism
here introduced.  As it has been shown in Figs.~\ref{sig1}--\ref{sig2}, it is possible
to reduce the angular velocity of the star by some 10--20$\%$ in $\sim
10$ years via emission of periodic GWs induced by r-mode
instabilities.  This reduction corresponds to an increase of the inner
pressure by roughly the same percentual amount.  As we have seen, when an
overpressure of order a few percent is reached, the star will reassess
forming a new region of MP. Therefore, we can expect a few periods of
GW bursts activity in 10 years. It is interesting to remark that in
this model the mechanism of emission of GWs is similar to the one
invoked in connection with the so-called soft-gamma repeaters
(see \cite{deFreitasPacheco:1998nn}). In particular, in both models the
burst of GWs is due to a starquake activity. The main difference
between the two models is that, in the soft-gamma repeaters case, the
origin of the sudden change of the structure of the star is related to
the maximum shear stress that the crust can bear before cracking.  In
the model proposed in the present paper, the role played by the
rigidity of the crust is now played by the surface tension at the
interface between hadronic and quark matter.  As we will show, it is
possible in our model to obtain more violent oscillations of the star
and therefore more energetic GW bursts. This is not surprising since
the physics underlying the soft-gamma repeaters model is based on a
metastability related to the typical energy scale of atomic-nuclear
physics, while in our model the energy scale is related to hadronic
physics.

%%%%%%%%%%%%%%%%%%%%%%%%%%%%%%%%%%%%%%%%%%%%%%%%%%%%%%%%%%%%%%%%%%%%%%%%%%%%%%%%%%%%%%%%%%%%%%
\begin{figure}[b]
\begin{center}
\includegraphics[scale=0.3]{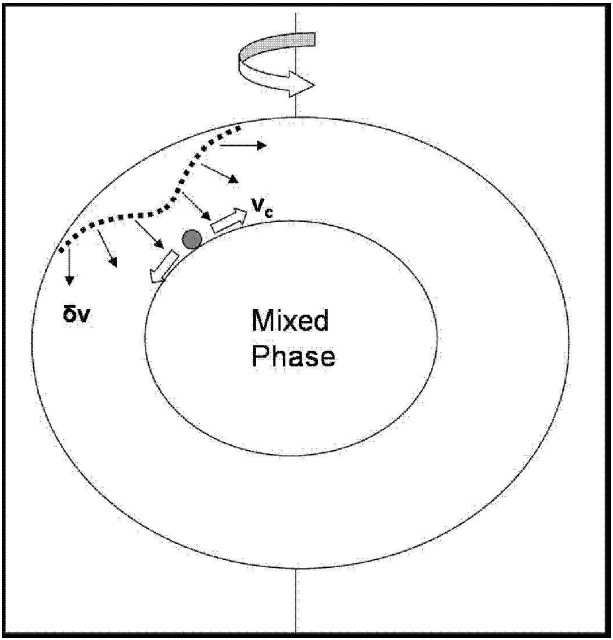}
\parbox{8cm}{
\caption {Picture of the mini-collapse process.
The nucleation site of the new drop of MP is indicated by the
small gray circle, $\vec v _c$ indicates the velocity of the
compression wave and $\delta \vec v$ is the velocity of the infalling
upper layers.
\label{disegno}}
}
\end{center}
\end{figure}
%%%%%%%%%%%%%%%%%%%%%%%%%%%%%%%%%%%%%%%%%%%%%%%%%%%%%%%%%%%%%%%%%%%%%%%%%%%%%%%%%%%%%%%%%%%%%%%%

In the model here described the emission of bursts of GWs is due to
the collapse of the region involved in the formation of new MP. It is
not easy to estimate the exact size of the interested region. Once a
drop of quark matter develops in the zone where MP is being produced,
the neighboring areas are probably triggered so that a relatively large
fraction of the metastable layer will collapse.  We can imagine that,
after the mini-collapse, a compression wave develops and propagates from the
site in which the new quark matter drop formed to the neighboring
regions as shown in the picture of Fig.~\ref{disegno}. 
The compression wave corresponds to a temporary overpressure that
adds on the already existing pressure inside the metastable matter. To
investigate if the propagation of the compression wave is able to trigger
the formation of new structures of quark matter, we compare two
timescales. The first timescale $t_\mathrm{trans}$ corresponds to the
duration of the overpressure $\delta p$, due to the compression wave, on a
specific point inside the metastable layer. It can be computed from
the velocity of the compression wave $v_c$ (which is the velocity of sound and it should be a fraction of
the velocity of light) and the size of the compression wave itself $\Delta$,
which should be of the same order as the radius of the newly formed
quark matter drop whose formation originated the wave,
i.e. $\Delta\sim$ 10 fm.  In this way we get $t_\mathrm{trans}\sim
10^{-20}$ s, where only an estimate of the order of magnitude is
needed.  This timescale must be compared with the nucleation time
$t_\mathrm{nucl}$ that corresponds to the value of overpressure
obtained adding the already existing overpressure in the metastable
layer $\Delta p$ to the overpressure due to the compression wave $\delta
p$. Assuming $\delta p\sim \Delta p$ and using the standard formalism
we obtain $t_\mathrm{nucl}\lesssim t_\mathrm{trans}$. We conclude
therefore that the conversion process can indeed propagate inside the
metastable layer with a velocity marginally lower than $v_c$.

Within our model, it is not possible to give a precise estimate of the
size of the region collapsing in a timescale of order 1 s or
smaller. The extension of this region can vary from the whole
metastable layer to a relatively small portion of the star. For
instance, it is possible that the conversion will propagate only in
the area in which new drops of QM are not too far from each other.
The randomness of the size of the collapsing region is typical of
quake phenomena. Once a fraction of the metastable layer has
collapsed, the other parts will presumably follow the same fate in a
timescale much shorter than the time needed to reach the critical
value of the overpressure. In other terms, we expect to have a few
bursts taking place in a relatively short period, while a much longer
delay (of order years) separates the phases of quake activity.  This
feature is similar to the temporal distribution observed in soft-gamma
repeaters (\cite{Cheng:1995}) and interpreted there as due to starquake
activity (\cite{deFreitasPacheco:1998nn}). Interestingly, the temporal
clustering of the events seems also to be a characteristic feature of
the GW bursts emission (\cite{Coccia:2004gw}).  A detailed analysis on
the temporal structure of the GW bursters has been made in
Ref.~(\cite{Dubath:2004vv}) where also the detection strategies of these
type of signals are discussed.

\subsection{Energy of the GW bursts}

We will now try to estimate the energy of the GW bursts.
It is not easy to provide a realistic approximation to this quantity 
since it would require a detailed analysis of the dynamics of the micro-collapse.
In this Section we will first present a rough estimate based essentially on dimensional analysis
and, after, a more quantitative discussion which will provide a lower limit to the energy released.

\subsubsection{Order of magnitude estimate}
The order of magnitude of the energy released can be estimated 
from the equation:
\be
E_{GW}=M \left({\Delta R/R}\right)^2\, ,
\ee
where $M$ is the mass in quadrupole motion and $\Delta R$ is the amplitude of the oscillation.
In our model, the entire region of the star above the layer where the conversion occurs
would participate to the quadrupolar motion, since the external regions have to readjust
immediately after the new layer of MP is formed. Therefore, we can assume that the 
mass in quadrupolar oscillation is of the same order of the total mass of the star.  
Concerning $\Delta R$, it is of the order of the 
shrinking of the radius of the star due to formation of a new layer of MP. 
Using the toy model results, we can estimate that for each mini-collapse the variation of 
the radius of the star is of order 20--30 m, which corresponds to 
$\Delta R/R\sim 2\div 3 \times 10^{-3}$. We can now put an upper limit
to the energy released in GWs. Assuming that the oscillation induced by the
phase transition is completely quadrupolar, the energy released would be of order
$E_{GW}\sim (0.5\div 1)\times 10^{-5} M_\odot$. 
In the next Section we will study how the oscillation decomposes into the normal
modes of the star and we will find that only a a few percent of the total energy
is associated with quadrupolar oscillations.
On the other hand, a more realistic model
should give larger variations of the radius for a same value of $\Delta\Omega/\Omega$.
We can expect that in a more realistic calculation the upper limit to the energy released in GWs can be up
to an order of magnitude larger, approaching $E_{GW}\sim 10^{-4} M_\odot$.

\subsubsection{Decomposition of the perturbation into nonradial modes}

In this Section we try to give a more quantitative estimate of the
amplitude of the GW, by explicitly performing the decomposition of the
perturbation into the normal nonradial modes of the star.  In the
following, we will first provide a ``model'' of the perturbation of
the star's fluid induced by the conversion process. Then, after
computing the spectrum of non-radial modes of the star, we decompose
the perturbation on the basis of the normal oscillation modes,
$\eta_{nl}$. In particular, we can compute the f-mode component
$\eta_{f2}$ of the perturbation and the corresponding GWs
emission. This is important because the excitation of
f-mode is the most efficient way for producing GWs.

In order to model the perturbation of the pressure of the star during
the micro-collapse, we resort again to the toy model and we assume
that the perturbation corresponds to the overpressure already
calculated in Sec.\ref{sectoy}, since immediately after the conversion
the overpressure reduces to zero.  In Sec.\ref{sectoy} we computed the
overpressure $\Delta P/P$ as a function of the effective
radius $a$, for fixed values of $\Omega$ and $\Delta
\Omega/\Omega$.  Inverting the relation given by Eq.~(\ref{radeff}),
we can write the overpressure as a function of $r$ and $\theta$. For
small values of the deformation parameter $\epsilon$ the perturbation can be approximated by:

\be
\frac{\Delta P}{\rho}\left(a(r,\theta)\right) = \frac{\Delta P}{\rho}\arrowvert_{\epsilon=0} 
+ r \frac{\mathrm{d} (\Delta P/\rho)}{\mathrm{d} r}\, \epsilon \,P_2(\cos(\theta)) \, .\label{perturb}
\ee

This corresponds to a decomposition of the perturbation into a radial
term, which will not contribute to the emission of GWs, and a
non-radial quadrupolar term, which will be further decomposed using
the normal non-radial modes basis.  In the following, while we compute
the nonradial modes within a realistic model for the star, concerning
the form of the perturbation $\Delta P/\rho$ we assume that it
is given by the expression obtained within the toy model. This crude
approximation can be overcome by computing the structure of a rotating
star using a realistic EOS (\cite{workinprogr}).  A generic perturbation
can be expanded as:

\be
\frac{\Delta P}{\rho}(r,\theta,\phi) = \sum_{n,l,m}c_{nlm}\eta_{nl}(r)Y_{l,m}(\theta,\phi)   \, ,\label{esp}  
\ee
where $\eta_{nl}(r)$ are the eigenfunctions of the non-radial oscillation equation
(see Ref.~(\cite{Unno:1989})).
The index $n$ can indicate the fundamental
$f$ mode or the $p$ and $g$ modes.
For simplicity,
we have computed the spectrum of non-radial oscillations in the case of a non-rotating, 
Newtonian Hybrid Star, within the Cowling approximation and assuming that
the Brunt-Vaisala frequency is vanishing. In this simplified case, g-modes
have zero frequency, and the equation describing non-radial oscillations
reduces to a Sturm-Liouville form.

%%%%%%%%%%%%%%%%%%%%%%%%%%%%%%%%%%%%%%%%%%%%%%%%%%%%%%%%%%%%%%%%%%%%%%%%%%%%%%%%%%%%%%%%%%%%%%%%%%%%%%%%%5 
\begin{figure}[b]
\begin{center}
\includegraphics[scale=0.52]{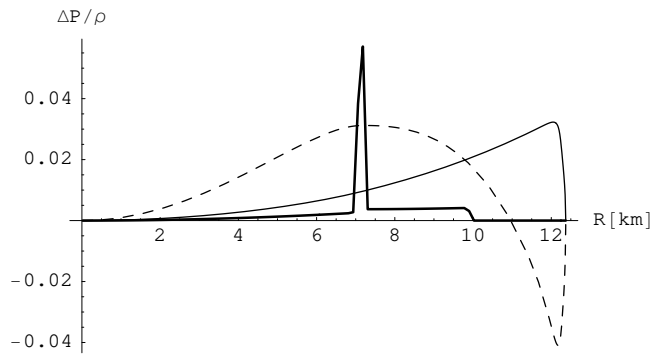}
\end{center}
\parbox{8cm}{
\caption{
\footnotesize The nonradial component of the perturbation $\Delta P/\rho$ (thick line) and
the f-mode (continuous line) and first p-mode eigenfunctions (for l=2), as functions of the radius of the star.}
\label{autofunzioni}}
\end{figure}
%%%%%%%%%%%%%%%%%%%%%%%%%%%%%%%%%%%%%%%%%%%%%%%%%%%%%%%%%%%%%%%%%%%%%%%%%%%%%%%%%%%%%%%%%%%%%%%%%%%%%%%%%%%%%%%%%%%%

In Fig.~\ref{autofunzioni} we display the quadrupolar component of the perturbation, defined in eq.~(\ref{perturb}), 
together with the $l=2$ eigenfunctions of the f-mode and of the first p-mode.
It is now possible to compute the f-mode component of the perturbation, which turns out to be
$c_{f2}\simeq 0.05$.

The power emitted in GWs by the excitation of the f-mode can be estimated using the formula (\cite{ipser-lindblom}):

\be
\label{dedt}
\left( \frac{dE}{dt} \right)_{\rm f2}= -\omega^{6} N_2 |\Delta D_{2 0}|^2
\ee

\noindent
where the general expression for $N_l$ reads
\be
N_l =  4 \pi \frac{G}{c^{2l+1}} \frac{(l+1)(l+2)}{l(l-1)[(2l+1)!!]^2},
\ee
$\omega$ is the frequency of the mode and $\Delta D_{lm}$ is the  
mass  multipole of the perturbed fluid. In our analysis we take 
into account only the quadrupolar contribution ($l=2$). 
%and $\Delta J_{lm}$ are the mass and current 
% multipoles of the perturbed fluid. In our analysis we will take into account only 
%quadrupolar contributions ($l=2$). 
The mass multipoles read:
\be
\Delta D_{lm} = \int { r^l \, \Delta \rho \, Y_{lm}^* d^3 x }\, ,\label{deltaM}
\ee 
where $\Delta \rho$ is the perturbation of the mass density.
The function $\Delta \rho$ can be expressed in terms of $\Delta P$, using the relation
\be
\Delta\rho=\frac{\rho}{\Gamma}\frac{\Delta P}{P}=\frac{\rho}{c_s^2}\frac{\Delta P}{\rho}\, 
\ee
where $\Gamma$ is the adiabatic index and $c_s$ is the sound velocity.
Finally, for $\Delta P/\rho$ we use the $l=2$, $f$-mode component 
obtained from eq.(\ref{esp}). Therefore the density multipole reads:
\be
(\Delta \rho )_{f2}= \frac{\rho}{c_s^2}\, c_{f2} \eta_{f2} Y_{2m}\, . 
\ee
Substituting this expression in eqs.~(\ref{dedt},\ref{deltaM})
we get an estimate of the power emitted in GWs due to the excitation of the f-mode:
\be
\left( \frac{dE}{dt} \right)_{f2}\sim 2.4 \times 10^{-7} \left (\frac{\omega/2\pi}{\mathrm{1 kHz}}\right )^6
M_\odot/\mathrm{s}\, .
\ee

In our estimate of the power of GWs emission we have considered up to now
only the contribution of the mass quadrupole,
but it is necessary to take into account also the
contribution associated with the current multipole, given by the expression:

\be
\delta J_{lm}= {2\over c}\sqrt{l\over l+1}
\int r^l (\delta \rho\, \vec{v}+\rho \,\delta\vec{v})\cdot
\vec{Y}^{B*}_{l\,m} d^3x, \label{flux}
\ee
where $\vec{Y}^{B*}_{l\,m}$ are the magnetic spherical harmonics. 
In Eq.~(\ref{flux}) two terms are present: one, $\vec v\, \delta\rho$, is
associated with the velocity of the fraction of the fluid
which undergoes the density fluctuation .
This term is suppressed respect to the one already
estimated, due to the presence of a factor $\vec v /c$ which is obviously
always smaller than one. 
The second term, $\rho\,\delta\vec v(\vec r)$, corresponds to the variation of velocity of
all fluid elements inside the star. In our model,
$\delta\vec v(\vec r)$ is associated with the rearrangement of the external parts
of the star, immediately after the formation of a new layer of MP.
It is very difficult to estimate $\delta\vec v(\vec r)$, and in particular 
its dependence on the position inside the star. We can
assume that all regions above the conversion layer
will move, as pictorially represented in Fig.~\ref{disegno}.
This term of the current multipole can be a more efficient source of GWs than 
the mass quadrupole, which is
relatively small because $\Delta P/\rho$ is non
vanishing only in a narrow region. On the other hand, the 
integral which defines the current multipole extends on a 
larger volume.
From simple order of magnitude estimate we found that its
contribution to the emitted power can be up to one order of magnitude larger than the one 
associated with the mass multipole.

\subsubsection{Frequency, time duration and amplitude of the GW bursts}

Concerning the value of the frequency of the emitted GW bursts,
in the simplified Newtonian scheme we have adopted, the
eigenfrequencies of the f-mode and the p-mode are
respectively of 3 and 8 kHz.  Calculations of these frequencies in
full General Relativity, give smaller values, typically of the order
of 2 kHz for the f-mode (\cite{Benhar:2004xg}). The corresponding duration
is $\tau_f\sim 0.18$ s (\cite{Benhar:2004xg}) and therefore the total energy emitted
by the f-mode is of order $E_f\sim 1.2\times 10^{-6}M_\odot$.

In the estimate of the f-mode frequency we have up to now neglected
the effect of rotation. It is known that in a rotating star the 
eigenfrequencies are shifted towards lower values (\cite{Ferrari:2003qu}).
In particular, the frequency of the f-mode reduces to a value
of order 1 kHz for fast rotating stars.

From the emitted power it is possible to estimate the initial amplitude $h_0$
of the GW using the equation (\cite{deFreitasPacheco:1998nn}):
\be
h_0=\left (-2 \frac{\mathrm{d}E}{\mathrm{d}t}\right )^{1/2}\frac{1}{D\,\omega_f}\, ,
\ee
where $D$ is the distance of the source.
We obtain therefore
\be
h_0\sim 3\times 10^{-21}\left (\frac{\omega/2\pi}{\mathrm{1 kHz}}\right )^2 \left({1 \mathrm{kpc}\over D}\right)\, .
\ee
The time dependence of the GW is given by
\be
h(t)=h_0 \exp ^{-(t/\tau_f-i\omega_f t)}\, ,
\ee
and the Fourier transform reads therefore:
\be
\tilde h (\omega)=\frac{h_0/\tau_f}{1/\tau_f\, ^2+(\omega-\omega_f)^2}\, .
\ee
If $\omega-\omega_f\sim 1000$ s$^{-1}$ then $\tilde h\sim (10^{-6}\, \mathrm{s}^2) h_0 /\tau_f$.
Instead if $\omega-\omega_f\sim 1/\tau_f$ then $\tilde h\sim h_0 \tau_f/2$.

It can be of interest to compare our results with the characteristics of the
signal detected by NAUTILUS and EXPLORER. The experimental
value of the Fourier transform of the amplitude is
$\tilde h\sim 2\times 10^{-21}$ s. Clearly our model can
approach this value only if the frequency $\omega_f$ is 
very near to the resonance frequency of the two gravitational bars,
which is of order of 1 kHz. Even in this case, to obtain in our model
an amplitude comparable to the experimental one it is necessary to
assume that the emitted power is significantly larger, maybe due to the
current multipole term and/or to the effect of using a more realistic model.

Finally, we want to stress that, even using the toy model,
the energy released in GW bursts is larger than the energy of the
bursts in other models of GW bursters, as e.g. soft gamma repeaters.
There, an energy  $\sim 10^{-10} - 10^{-9} M_\odot$ can be released, but only under
the assumption that all the
elastic stress is completely converted into GW bursts (\cite{deFreitasPacheco:1998nn}).

\section{Conclusions}

We would like finally to discuss the phenomenological
relevance of the proposed model. 
Concerning the stellar objects which can be possible
candidates for our model, if we assume that all neutron stars are
born with a large value of angular velocity, they will all enter the
instability window as described in the first scenario.  Taking a
neutron star production rate of order 0.02 per year in our galaxy, and
assuming the possibility to detect GW bursts up to a distance of order
1 kpc, the probability of finding an active burster in this region is
of order percent, if the total duration of the emission phase is of
order 50 years.  Moreover, if the possibility of ``recycling''
described in the second scenario is taken into account, the
probability can be larger. A precise estimate of the probability would
require a precise knowledge of the number of accreting millisecond
pulsars, what is not known at the moment.

An important feature of the model discussed here is that no neutrino
signal is expected in detectors as LVD (in the Gran Sasso National
Laboratory) or even in more massive detectors as Super Kamiokande. In
our model an energy $E_\mathrm{trans}\sim 10^{-4} M_\odot$ is
deposited inside the star, near the metastable layer, during the
transition.  The neutrinos produced by URCA processes have typical
energies of a few MeV and they will scatter many times inside the
star, degrading their energy before escaping. Therefore the emitted
neutrinos have energies below the threshold of neutrino detectors.

Finally, let us discuss possible signatures of the mechanism here
proposed.  Concerning the number of epochs during which bursts are
emitted, we can expect that it will decrease from a few active periods
in ten years to a few periods in the next hundred years, because the
angular velocity decreases by roughly the same amount in the first ten
years as in the following hundred years.  The amplitude of the signal
should also decrease because, as shown in Figs.~\ref{raggi} and
\ref{over}, when the angular velocity is lower, the variation of the
radius of the star is also smaller.  Both these features are related
to the draining of the bursts energy source which, in this model, is
ultimately the rotational energy of the star.

\section{Acknowledgments}

It is a pleasure to thank E.~Coccia, G.~Fiorentini and M.~Maggiore for very useful discussions.
A special thank to A.~Ortolan for suggesting the possible connection between sudden variations 
of the star structure and the emission of short GW bursts.

\bibliography{references}
\bibliographystyle{aa}
\end{document}